\documentstyle[twocolumn,aps,epsfig]{revtex}
\begin{document}
\draft
\title{Gauge Invariant Formulations of Dicke-Preparata Super-Radiant
Models}
\author{S. Somu and A. Widom}
\address{Physics Department, Northeastern University, Boston MA 02115}
\author{Y.N. Srivastava}
\address{Physics Department \& INFN, University of Perugia, Perugia Italy}
\maketitle

\begin{abstract}
In a gauge invariant formulation of the molecular electric 
dipole-photon interaction, the rigorous coupling is strictly 
linear in the photon creation and photon annihilation operators. 
The linear coupling allows for a super-radiant phase transition 
as in the Hepp-Lieb formulation. A previous notion of a 
quadratic-coupling ``no-go theorem'' for super-radiance is incorrect. 
Also incorrect is a previous assertion that the dipole-photon 
coupling has absolutely no effect on the thermal equations of state.  
These dubious assertions were based on incorrect canonical 
transformations which eliminated the electric field (and thereby  
eliminated the dipole-photon interaction) which is neither 
mathematically nor physically consistent. The correct form of the 
canonical transformations are given in this work which allows for the 
physical reality of super-radiant condensed matter phases. 
\end{abstract}  

\narrowtext

\section{Introduction} 

In an early paper on the interaction between radiation and molecular 
dipole moments \cite{1}, Dicke formulated a model which was later shown to 
exhibit a super-radiant phase transition\cite {2,3}. The notion that such  
phase transitions should exist in condensed matter systems has been 
investigated in a series of papers by Preparata and coworkers\cite {4,5,6}
and others\cite {7,8,9}. Different workers have come to somewhat different 
conclusions concerning super-radiant phase transitions
\cite {10,11,12,13,14,15,16,17,18,19}. Some doubt has 
been expressed\cite {20,21,22,23,24} concerning the physical 
laboratory reality of super-radiant phase transition. The mathematical
 issues are as follows: 
(i) It appears,{\em at first glance}, that quadratic terms (in photon
 creation and annihilation operators) enter into the model via quadratic 
terms in the vector potential \begin{math} {\bf A}  \end{math}. (ii) The 
quadratic terms in the ``corrected Dicke model'' appear to destroy the 
super-radiant phase transition. 

Our purpose is to show that if the dipole-field interaction 
is treated in a {\em gauge invariant manner}\cite{25,26,27}
 then the interaction is {\em strictly linear} in the electric field 
\begin{math}{\bf E} \end{math}. Thus, quadratic terms are {\em absent} 
for purely electric dipole-photon interactions [28]. These considerations 
render likely the physical reality of condensed matter super-radiant 
phase transitions.

To see what is involved, let us first consider a single molecule with an 
electric dipole moment \begin{math} {\bf \mu} \end{math} given by  
\begin{equation}
{\bf \mu}=\sum_{a=1}^N q_a {\bf r}_a,
\end{equation}
where \begin{math} q_a  \end{math} and \begin{math} {\bf r}_a  \end{math}
denote respectively, the charge and position of the 
\begin{math} a^{th}  \end{math} particle. Suppose that the dipole moment 
interacts with a {\em classical} electric field which is uniform in 
space, but not in time 
\begin{equation}
{\bf E}(t)=-{1\over c}\left({d{\bf A}(t)\over dt}\right).
\end{equation}
The molecule will have a time dependent Hamiltonian of the form 
\begin{equation}
H_{mol}(t)=\sum_{a=1}^N K_a\left({\bf p}_a-{q_a\over c}{\bf A}(t)\right)
+V({\bf r}_1,...,{\bf r}_N),
\end{equation}
where \begin{math} K_a({\bf p}_a)=({\bf p}_a^2/2m_a) \end{math} is 
the kinetic energy of \begin{math} a^{th} \end{math} particle in the 
molecule and \begin{math} V({\bf r}_1,...,{\bf r}_N)   \end{math} is 
the internal Coulomb energy of the molecule. Employing the unitary operator 
\begin{equation}
U(t)=e^{i\left({\bf \mu \cdot A}(t)/\hbar c \right)}
\end{equation}
in the time dependent canonical transformation 
\begin{equation}
{\cal H}_{mol}(t)=U^\dagger (t)H_{mol}(t)U(t)-i\hbar U^\dagger (t)
{dU(t)\over dt}
\end{equation}
yields 
\begin{equation}
{\cal H}_{mol}(t)=\sum_{a=1}^N K_a({\bf p}_a)+V({\bf r}_1,...,{\bf r}_N)
-{\bf \mu \cdot E}(t).
\end{equation}
For completeness of presentation, a derivation of Eq.(5) is given in 
Appendix A.

Note: For classical electric fields of the form in Eq.(2), the 
Hamiltonians in Eqs.(3) and (6) are {\em rigorously} equivalent. 
Furthermore, the Hamiltonian in Eq.(3) depends on linear and quadratic
terms in the vector potential \begin{math} {\bf A}(t) \end{math}. On the 
other hand, the Hamiltonian in Eq.(6) depends only on the electric field 
\begin{math} {\bf E}(t) \end{math} and in a strictly {\em gauge invariant} 
and {\em linear} fashion. 

In the work which follows, both the charged particles and the
electromagnetic field will be treated using quantum mechanics.  
In Sec.II we discuss a single molecule in the presence of the 
quantized electromagnetic field. It is proved for the gauge 
invariant description of the dipole interactions that the 
coupling between the electromagnetic field and the molecule is 
linear in photon creation and annihilation operators  The results 
for \begin{math} {\cal N}  \end{math} molecules are discussed in 
Sec.III. The resulting Hamiltonian has four terms: 
(i) a sum of single molecule Hamiltonians, (ii) a sum of screened 
Coulomb interaction potentials between neighboring molecules, 
(iii) the radiation field energy and (most importantly) 
(iv) a {\em strictly linear coupling} between the dipole moments 
and the electric field. Dicke-Preparata models are defined and 
explored in Sec.IV. In the concluding Sec.V, the importance of the 
super-radiant phase transition associated with linear couplings 
will be discussed. The errors by some previous workers who falsely 
``voided'' the super-radiant phase transition will be discussed in 
detail. 

\section{Photon-Dipole Interactions}

Consider, as in Sec.I, a single molecule having a total number of 
\begin{math} N \end{math} charged particles with 
an electric dipole moment \begin{math} {\bf \mu} \end{math} 
as in Eq.(1). The dipole is considered to interact with  
a quantized electromagnetic field (photons) 
\begin{math} {\bf A} \end{math} and/or \begin{math} {\bf E} \end{math} 
which is again uniform in space. The Hamiltonian for the photons is 
given by 
\begin{equation}
H_{rad}=\sum_k \hbar \omega_k a^\dagger _{k} a_k,   
\end{equation}
where 
\begin{equation}
\left[ a_{k^\prime }, a^\dagger _{k} \right] = \delta_{kk^\prime }.
\end{equation}
Furthermore, for a ``quantization box'' of volume 
\begin{math} \Omega  \end{math}, 
\begin{equation}
{\bf A}=c\sqrt{2 \pi \hbar \over \Omega }\sum_k {1\over \sqrt{\omega_k}}
({\bf e}_k a_k+{\bf e}^*_ka^\dagger _k ),
\end{equation}
and 
\begin{equation}
{\bf E}=i\sqrt{2 \pi \hbar \over \Omega }
\sum_k{\sqrt{\omega_k }}({\bf e}_k a_k-{\bf e}^*_k a^\dagger _k ).
\end{equation}
The matter Hamiltonian for a single molecule in the electric 
dipole limit is then  
\begin{equation}
H_{mol}({\bf A})
=\sum_{a=1}^N K_a\left({\bf p}_a-{q_a\over c}{\bf A}\right) 
+ V({\bf r}_1,...{\bf r}_N).  
\end{equation}

In total, the time independent Hamiltonian for the molecule 
with dipole-quantized electromagnetic field interactions reads 
\begin{equation}
H=H_{rad}+H_{mol}({\bf A}),
\end{equation}
where \begin{math} H_{rad}  \end{math} is defined in Eq.(7), and 
\begin{math} H_{mol}({\bf A})  \end{math} is defined in 
Eqs.(9) and (11). In the quantum electrodynamic model, Eq.(2) is 
replaced by the operator equation 
\begin{equation}
{\bf E}=-\left({i\over \hbar c}\right)\left[H,{\bf A}\right]
\end{equation}
leading to Eq.(10).

Employing the unitary operator
\begin{equation}
S=e^{i({\bf \mu \cdot A}/ \hbar c )}
\end{equation}
in the time-independent canonical transformation 
\begin{equation}
{\cal H}=S ^\dagger H S,
\end{equation}
yields 
\begin{equation}
{\cal H}=H_{rad}+H_{mol}({\bf A}=0)-{\bf\mu} \cdot{\bf E}+ W_{mol}
\end {equation} 
where the ``self energy'' contribution to the molecule 
is given by 
\begin{equation}
W_{mol}={2 \pi \over \Omega}\sum_k ({\bf e}_k{\bf \cdot \mu })
({\bf e}^*_k{\bf \cdot \mu }).
\end{equation}
The derivation of Eq.(16) is given in Appendix B.

Introducing the boson operators \begin{math} b_k=ia_k \end{math} 
and the Hermitian conjugate 
\begin{math} b^\dagger _k=-ia^\dagger _k  \end{math} yields 
our final Hamiltonian 
$$
{\cal H}=\sum_k \hbar \omega_k  b^\dagger _k b_k
+\sum_{a=1}^N K_a ({\bf p}_a) + V({\bf r}_1,...{\bf r}_N)
$$
\begin{equation}
+\left({2 \pi \over \Omega}\right)\sum_k ({\bf e}_k{\bf \cdot \mu })
({\bf e}^*_k{\bf \cdot \mu })-{\bf \mu \cdot E}.
\end{equation}
where 
\begin{equation}
{\bf E}=\sqrt{2 \pi \hbar \over \Omega }
\sum_k{\sqrt{\omega_k }}({\bf e}_k b_k+{\bf e}^*_k b^\dagger _k ),
\end{equation}
and 
\begin{equation}
\left[ b_{k^\prime }, b^\dagger _{k} \right] = \delta_{kk^\prime }.
\end{equation}
The central result of this section is the following:
\medskip 
\par \noindent 
{\bf Theorem:} For the electric dipole-electromagnetic field interaction,  
the resulting Hamiltonian is that of free photons plus interaction terms 
linear in the photon creation and annihilation operators. {\em Quadratic 
interaction terms rigorously vanish.}

\medskip 
\par \noindent 
{\bf Proof:} See Eqs.(18) and (19), which were derived solely on the basis 
of the dipole interaction.
\medskip
\par \noindent
Let us now consider the case of \begin{math} {\cal N}  \end{math}
molecules, 
each of which interact with the quantized electromagnetic field via the 
electric dipole moment.

\section{Many Molecules}

We consider the case of many molecules with electric dipole moments 
interacting with electromagnetic field modes whose minimum 
wavelength \begin{math} \lambda_{min}  \end{math} is large on 
the scale of the molecular size \begin{math} L \end{math}; i.e. 
\begin{math} \lambda_{min}>>L \end{math} as shown in Fig.1 below.  

\begin{figure}[htbp]
\begin{center}
\mbox{\epsfig{file=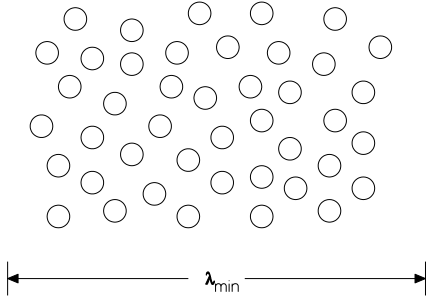,height=60mm}}
\caption{Shown are many molecules of small size $L$ when compared with the 
minimum photon wavelength $\lambda_{min}$.}
\label{topfig1}
\end{center}
\end{figure}

The Hamiltonian for \begin{math} {\cal N}  \end{math} 
molecules plus the electromagnetic field reads 
\begin{equation}
H=H_{rad}+\sum_{j=1}^{\cal N}H_{mol;j}({\bf A})+
\sum_{1\le j<i\le {\cal N}}V_{ij}.
\end{equation}
The radiation Hamiltonian in the Coulomb gauge 
\begin{equation}
div{\bf A}({\bf r})=0
\end{equation}
is given by 
\begin{equation}
H_{rad}={1\over 8\pi }\int \left({\bf E}^2+{\bf B}^2\right)d^3{\bf r}.
\end{equation}
The internal \begin{math} j^{th} \end{math} molecular Hamiltonian is 
given by 
$$
H_{mol;j}({\bf A})=
$$
\begin{equation}
\sum_{a=1}^N K_a\left({\bf p}_{(a;j)}-{q_a\over c}{\bf A}({\bf
R}_j)\right)
+V({\bf r}_{(1;j)},...,{\bf r}_{(N;j)}),
\end{equation}
where \begin{math} {\bf R}_j  \end{math} is the central position of 
the \begin{math} j^{th}  \end{math} molecule. 

The internal Hamiltonian of each molecule contains the internal 
Coulomb potentials. In the last term on the right hand side of Eq.(21), 
\begin{math} V_{ij} \end{math} describes the Coulomb interactions between 
two different molecules. The long ranged part of this potential is of the 
conventional dipole-dipole interaction form 
\begin{equation}
V^{dipole}_{ij}= {\bf \mu}_i \cdot \left({
|{\bf R}_{ij}|^2 {\bf 1}-3{\bf R}_{ij}{\bf R}_{ij}\over |{\bf R}_{ij}|^5
}\right)\cdot {\bf \mu}_j 
\end{equation}
where \begin{math} {\bf R}_{ij}={\bf R}_i-{\bf R}_j \end{math}. 

The commutation relations of the quantum electrodynamic fields in the 
Coulomb gauge are given by 
\begin{equation}
\left[{\bf E}({\bf r}),{\bf A}({\bf r}^\prime )\right]=
4\pi i \hbar c {\bf \Delta }({\bf r}-{\bf r}^\prime )
\end{equation}
where the transverse delta function is defined as 
\begin{equation}
{\bf \Delta }({\bf r})=\int \left({\bf 1}-\hat{\bf k}\hat{\bf k}\right)
e^{i{\bf k\cdot r}}\left({d^3 {\bf k}\over (2\pi )^3}\right).
\end{equation}
Equivalently 
\begin{equation}
{\bf \Delta }({\bf r})=\int 
\left({|{\bf k}|^2{\bf 1}-{\bf k}{\bf k}\over |{\bf k}|^2}\right)
e^{i{\bf k\cdot r}}\left({d^3 {\bf k}\over (2\pi )^3}\right),
\end{equation}
so that 
\begin{equation}
{\bf \Delta }({\bf r})=\left({1\over 4\pi }\right)
\left({\bf \nabla }{\bf \nabla }-{\bf 1}\nabla^2 \right){1\over |{\bf
r}|}.
\end{equation}
Explicitly\cite{29} 
\begin{equation}
{\bf \Delta }({\bf r})=\left({2\over 3}\right){\bf 1}\delta ({\bf r})
+\left({1\over 4\pi }\right)
\left({3{\bf r}{\bf r}-|{\bf r}|^2{\bf 1}\over |{\bf r}|^5}\right).
\end{equation}
Eqs.(25) and (30) imply 
\begin{equation}
V^{dipole}_{ij}= -4\pi {\bf \mu }_i \cdot 
{\bf \Delta }({\bf R}_{ij})\cdot {\bf \mu }_j\ \ {\rm for}\ \ 
{\bf R}_{ij}\ne 0. 
\end{equation}

The transverse distribution function 
\begin{math} {\bf \Delta }({\bf r})  \end{math} removes the 
longitudinal part of a vector field. For example, if 
\begin{equation}
\tilde{\bf P}({\bf r})=
\sum_{j=1}^{\cal N}{\bf \mu }_j \delta ({\bf r}-{\bf R}_j)
\end{equation}
represents polarization, i.e.  the molecular dipole moment per unit 
volume, then 
\begin{equation}
{\bf P}({\bf r})=\int {\bf \Delta }({\bf r}-{\bf r}^\prime )
\cdot \tilde{\bf P}({\bf r}^\prime )d^3 {\bf r}^\prime
\end{equation}
represents the transverse part of the polarization.

Now, consider the unitary operator 
\begin{equation}
{\cal S}=\exp\left({i\over \hbar c}
\int {\bf A}({\bf r})\cdot  \tilde{\bf P}({\bf r})d^3 {\bf r}\right).
\end{equation}
Eq.(26), (33) and (34) imply  
\begin{equation}
{\cal S}^\dagger {\bf E}({\bf r}){\cal S}=
{\bf E}({\bf r})-4\pi {\bf P}({\bf r})
\end{equation}
so that the radiation Hamiltonian of Eq.(23) reads 
\begin{equation}
{\cal S}^\dagger H_{rad}{\cal S}={1\over 8\pi }
\int \big(|{\bf E}-4\pi {\bf P}|^2+|{\bf B}|^2\big)d^3{\bf r}.
\end{equation} 
Eqs.(17), (23), (31)-(33) and (36) imply that 
$$
{\cal S}^\dagger H_{rad}{\cal S}=H_{rad}-
\int {\bf E}\cdot {\bf P}d^3{\bf r}
$$
\begin{equation}
+\sum_{j=1}^{\cal N}W_{mol;j}\ \ 
-\sum_{1\le i<j\le {\cal N}}V^{dipole}_{ij}.
\end{equation} 
It is useful to define the ``screened'' (or short ranged) 
intermolecular Coulomb potential with the long ranged dipole-dipole 
potential subtracted
\begin{equation}
\tilde{V}_{ij}=V_{ij}-V^{dipole}_{ij}.
\end{equation}
Finally, the unitary transformation removes the vector potential 
from the internal degrees of freedom of a molecule as in the above 
Sec.II; i.e. in the dipole interaction limit  
\begin{equation}
{\cal S}^\dagger H_{mol;j}({\bf A}) {\cal S}=H_{mol;j}({\bf A}=0).
\end{equation}

The ``grand finale'' of the algebraic ceremony of this section is the 
assertion that total transformed Hamiltonian 
\begin{equation}
{\cal H}={\cal S}^\dagger H {\cal S}
\end{equation}
which for \begin{math} {\cal N} \end{math} molecules takes the form 
$$
{\cal H}={1\over 8\pi }\int \left({\bf E}^2+{\bf B}^2\right)d^3{\bf r}+
\sum_{1\le i<j\le {\cal N}}\tilde{V}_{ij}
$$
\begin{equation}
+\sum_{j=1}^{\cal N} {\cal H}_{mol;j}-
\int {\bf E}\cdot {\bf P}d^3{\bf r},
\end{equation}
where the internal molecular Hamiltonian  
\begin{equation}
{\cal H}_{mol;j}=H_{mol;j}({\bf A}=0)+W_{mol;j}
\end{equation}
is independent of the transverse \begin{math} {\bf A}  \end{math}.
The central Eq.(41) of this section follows from Eqs.(17), (24), 
(37)-(39) and (40).

\section{Dicke-Preparata Model}

The Hamiltonian in Eq.(41) has four parts: (i) a sum of intramolecular 
Hamiltonians describing the internal kinetic and Coulomb energies of 
each molecule, (ii) a sum of screened (short ranged) intermolecular 
Coulomb potentials between neighboring molecules, (iii) the radiation 
field energy and (iv) a linear coupling between the dipole moment per 
unit volume and the electric field. Within the dipole approximation, 
Eq.(41) is exact but not yet amenable to rigorous mathematical solution. 
For the purpose of obtaining reasonable answers for (say) the free energy, 
further simplifications must be made.  

For the Dicke-Preparata model, the approximations are as follows: 
(i) The sum of intramolecular Hamiltonians is modeled by a sum of 
``two-level molecules'', each described by Pauli matrices 
\begin{math} (\sigma_{x;j},\sigma_{y;j},\sigma_{z;j}) \end{math}; 
\begin{equation}
\sum_{j=1}^{\cal N} {\cal H}_{mol;j}\Rightarrow 
{\varepsilon \over 2}\sum_{j=1}^{\cal N} \sigma_{z;j}.
\end{equation} 
(ii) The sum of screened intermolecular Coulomb potentials between 
neighboring molecules is neglected
\begin{equation}
\sum_{1\le i<j\le {\cal N}}\tilde{V}_{ij}\Rightarrow 0.
\end{equation}
(iii) The radiation field energy is modeled by a single photon 
oscillator mode 
\begin{equation}
{1\over 8\pi }\int \left({\bf E}^2+{\bf B}^2\right)d^3{\bf r}
\Rightarrow \hbar \omega_0 b^\dagger b.
\end{equation}
(iv) The linear coupling between the dipole moment per 
unit volume and the electric field is modeled by the {\em resonant} 
photon creation and annihilation terms 
\begin{equation}
-\int {\bf E}\cdot {\bf P}d^3{\bf r}\Rightarrow 
\tilde{\mu }\sqrt{2\pi \hbar \omega_0\over \Omega }\ 
\sum_{j=1}^{\cal N}(b\sigma_{+;j}+b^\dagger \sigma_{-;j}),
\end{equation}
where \begin{math} 2\sigma_{\pm ;j}=\sigma_{x ;j}\pm
i\sigma_{y;j}\end{math} 
and \begin{math} \tilde{\mu }  \end{math} is the excitation matrix element 
of the electric dipole operator of a single molecule. Putting the model 
replacement Eqs.(43)-(46) into Eq.(41) yields the Dicke-Preparata model 
\begin{equation}
{\cal H}\Rightarrow {\cal H}_{DP}
\end{equation}
where 
\begin{equation}
{\cal H}_{DP}=\varepsilon S_z+\hbar \omega_0 b^\dagger b 
+\tilde{\mu }\sqrt{2\pi \hbar \omega_0\over \Omega }
\ \left(bS_+ +b^\dagger S_-\right),
\end{equation}
\begin{equation}
S_z={1\over 2}\sum_{j=1}^{\cal N}\sigma_{z;j},
\end{equation}
and 
\begin{equation}
S_{\pm }={1\over 2}\sum_{j=1}^{\cal N}
(\sigma_{x;j}\pm i\sigma_{y;j}).
\end{equation}

The Dicke-Preparata model may be written in terms of the oscillator 
coordinate \begin{math} Q  \end{math} and momentum 
\begin{math} P \end{math} using 
\begin{equation}
b={(P -i\omega_0 Q)\over \sqrt{2\hbar \omega_0}},\ \ \ 
b^\dagger = {(P +i\omega_0 Q)\over \sqrt{2\hbar \omega_0}},
\end{equation}
yielding 
\begin{equation}
{\cal H}^\prime_{DP}=\left({P^2+\omega_0^2Q^2\over 2}\right)
+{\bf h \cdot S}
\end{equation}
where the vector \begin{math} {\bf h} \end{math} is given by 
\begin{equation}
{\bf h}=\tilde{\mu }\sqrt{4\pi \over \Omega }
\left(P{\bf i} +\omega_0 Q{\bf j}\right)+\varepsilon {\bf k}.
\end{equation}

The free energy \begin{math} {\cal F}  \end{math} of the model is 
determined by 
\begin{equation}
e^{-{\cal F}/k_B T}=
Tr\ e^{-{\cal H}^\prime_{DP}/k_B T}.
\end{equation}
Here, the trace over the oscillator degree of freedom may be taken as 
being classical with virtually zero error in the limit 
\begin{math} {\cal N}\to \infty  \end{math}; i.e.  
\begin{equation}
e^{-{\cal F}/k_BT}=
\int_{-\infty}^\infty \int_{-\infty}^\infty 
tr\ e^{-{\cal H}^\prime_{DP}/k_BT}
\left({dPdQ\over 2\pi \hbar}\right).
\end{equation}
Since 
\begin{equation}
tr\ e^{-{\bf h\cdot S}/k_B T}=
\left\{2\ \cosh\left({|{\bf h}|\over 2k_B T}\right) \right\}^{\cal N},
\end{equation}
and since Eq.(53) implies 
\begin{equation}
|{\bf h}|^2=\varepsilon^2 +
\left({4\pi \tilde{\mu }^2\over \Omega }\right)(P^2+\omega_0^2 Q^2), 
\end{equation} 
it follows from Eqs.(52), (53), and (55)-(57) that  
\begin{equation}
e^{-{\cal F}/k_B T}=
\int_{-\infty}^\infty \int_{-\infty}^\infty e^{-{\cal G}/k_BT}
\left({dPdQ\over 2\pi \hbar}\right),
\end{equation}
where
\begin{equation}
{{\cal G}\over {\cal N}}=w-k_B T\ ln\left(2 \cosh
\left\{ 
{\sqrt{\varepsilon^2+8\pi \tilde{\mu}^2nw}\over 2k_B T}  
\right\}\right),
\end{equation}
the oscillator energy per molecule is
\begin{equation}
w=\left( {P^2+\omega_0^2 Q^2 \over 2 {\cal N} }\right),
\end{equation}
and the number of molecules per unit volume 
\begin{math} n \end{math} is  
\begin{equation}
n=({\cal N}/\Omega ).
\end{equation}

In the thermodynamic limit of Eqs.(58)-(61), 
the minimum free energy principle is that 
\begin{equation}
\lim_{{\cal N}\to \infty }({\cal F}/{\cal N})=
\inf_{(0<w<\infty ) }\ f(w,n,T),
\end{equation}  
where 
\begin{equation}
f(w,n,T)=w-k_B T\ ln\left(2 \cosh \chi(w,n,T)\right),
\end{equation}
and
\begin{equation}
\chi(w,n,T)={\sqrt{\varepsilon^2+8\pi \tilde{\mu}^2nw}\over 2k_B T}\ .
\end{equation}

Let \begin{math} \bar{w}(n,T)\ge 0 \end{math} represent the mean oscillator 
energy per molecule as a function of the molecular density 
\begin{math} n  \end{math} and the temperature \begin{math} T \end{math}.
If \begin{math} \bar{w}>0 \end{math}, then the system is in a super-radiant 
phase. If \begin{math} \bar{w}=0 \end{math}, then the system is in a 
normal (incoherent radiation) phase. The free energy per molecule  
\begin{math} f(w,n,T)  \end{math} is at a minimum for  
\begin{math} w=\bar{w} \end{math}. Setting 
\begin{math} (\partial f/\partial w)_{nT}=0 \end{math} yields the following 
implicit function equation for \begin{math} \bar{w} \end{math}:
\begin{equation}
\left({k_BT \chi(\bar{w},n,T)\over \pi \tilde{\mu }^2n}\right)=
\tanh \chi(\bar{w},n,T).
\end{equation} 

The phase diagram in the \begin{math} (n,T) \end{math} plane follows 
from setting  \begin{math} \bar{w}(n,T_c)=0 \end{math} and solving for 
the critical temperature \begin{math} T_c (n) \end{math} as a function of 
density. The super-radiant phase exists for 
\begin{math} T<T_c (n) \end{math} and the normal phase exists for 
\begin{math} T>T_c (n) \end{math}. The critical temperature 
\begin{math} T_c (n\le n_c)=0 \end{math} where the critical density 
\begin{equation}
n_c=\left({\varepsilon \over 2\pi \tilde{\mu }^2}\right).
\end{equation} 
Analytically,
\begin{equation}
k_B T_c(n>n_c)=\left\{
{\varepsilon \over ln\big((n+n_c)/(n-n_c)\big)}
\right\}.
\end{equation}
A plot of the phase diagram is shown below.

\begin{figure}[htbp]
\begin{center}
\mbox{\epsfig{file=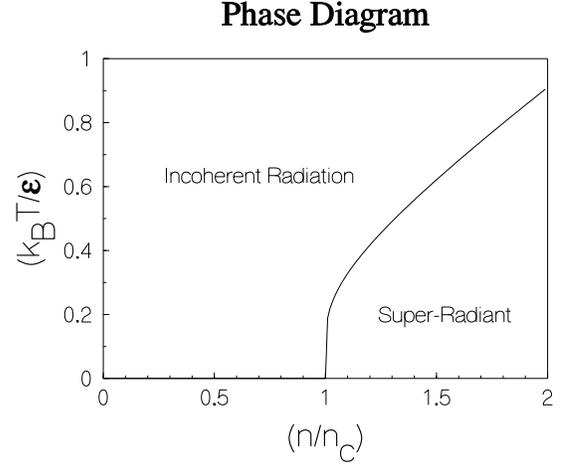,height=80mm}}
\caption{Shown is the phase diagram in the $(n,T)$-plane, 
where $n_c$ is defined in Eq.(66). The two 
phases are separated by the critical temperature curve 
$(k_BT_c(n)/\varepsilon )$.}
\label{topfig2}
\end{center}
\end{figure}

\section{Conclusions}

The doubts about the physical reality of the super-radiant phases were 
discussed in great detail by I. Bialynicki-Birula  and K. Rzazewski\cite{23}, 
who came to the remarkable (and dubious) conclusion that the electric 
dipole-photon interaction could be gauged away. This incorrect conclusion 
was based on using a vector potential which is independent of both space 
and time. It is no wonder that such a pure ``gauge field'' had zero 
effect on the thermodynamic equations of state. 

In reality, the transverse electric field in the Coulomb gauge should 
be calculated using either
\begin{equation}
{\bf E}=-{1\over c}{\partial {\bf A}\over \partial t },\ \ 
{\rm (Heisenberg\ or\ Classical\ Picture)}, 
\end{equation}    
or 
\begin{equation}
{\bf E}=-\left({i\over \hbar c }\right)\left[ H, {\bf A} \right], \ \ 
{\rm (Schrodinger\ Picture)}.
\end{equation}
The {\em classical canonical transformation} Eq.(12) 
of I. Bialynicki-Birula  and K. Rzazewski\cite{23} (in the dipole 
approximation) should have been properly performed as in our Eqs.(5) 
and (6). The last term on the right hand side of both our Eqs.(5) and (6) 
were incorrectly omitted in Eq.(12) of I. Bialynicki-Birula  and 
K. Rzazewski\cite{23}. This incorrect classical canonical transformation 
eliminated the electric field \begin{math} {\bf E}  \end{math} and 
thereby the electric dipole-photon interaction.

A proper procedure is to employ a fully quantum electrodynamic 
framework to obtain a non-trivial electric field 
\begin{math} {\bf E}  \end{math} {\em before} going to the classical 
electrodynamic oscillator trace evaluation. From this rigorous 
Schr\"odinger picture point of view, Eq.(12) of I. Bialynicki-Birula  
and K. Rzazewski\cite{23} is still {\em incorrect} because Eqs.(B2-B3) 
in our Appendix B  Eqs.(B2-B3) were not taken into account in that 
work\cite{23}. 

In any or all pictures, the electric dipole-photon interaction {\em 
cannot} be described by gauging away the electric field 
\begin{math} {\bf E} \end{math}. When the Hamiltonian is expressed in 
terms of \begin{math} {\bf E} \end{math} and 
\begin{math} {\bf B} \end{math}, it is then properly gauge invariant and 
non-trivial changes {\it very surely} appear in the thermal equations 
of state. For example, the one-photon loop corrections to the equations 
of state have long been known to produce Casimir-van der Waals forces 
in condensed matter. Such forces cannot possibly be gauged away and 
I. Bialynicki-Birula  and K. Rzazewski\cite{23} are thereby in error.
A further error is found in K. Rzazewski, K. Wodkiewicz and 
W. Zakowicz\cite{20}, where it is asserted that the super-radiant phase 
transition is null and void due to quadratic coupling in the photon 
creation and annihilation operators. Had the canonical transformation 
been carried out correctly, as in the present work, then it would have 
been noted that the coupling between the electric dipole moment and 
the photons in the gauge invariant Eq.(41) is {\em strictly linear}. 
The super-radiant phase transition remains intact.  

Preparata has recently proposed several examples of super-radiant phases 
in a recent review\cite{30}. Water is perhaps the most interesting 
example\cite{4,31}. In future work it would appear interesting to relate 
the super-radiant transition to an instability of the incoherent radiation 
phase which produces the conventional theoretical and experimental QED 
Casimir effects. QED effects at the one photon loop level have long  
proved their importance in the understanding of the long ranged 
intermolecular forces\cite{29}.

\bigskip 
\centerline{\bf MATHEMATICAL APPENDIX A}
\bigskip

Starting from the Schr\"odinger equation 
$$
i\hbar \left({ d\left|\phi (t)\right> \over dt}\right)=
H(t)\left|\phi (t)\right>,
\eqno(A1)
$$
one seeks a solution of the form 
$$
\left|\phi (t)\right>=U(t)\left|\psi (t)\right>,
\eqno(A2)
$$
where $U^\dagger(t)=U^{-1}(t)$. Putting Eq.(A2) into Eq.(A1) yields 
$$
i\hbar U(t)\left({d\left|\psi (t)\right> \over dt}\right)+
i\hbar \left({dU(t)\over dt}\right)\left|\psi (t)\right>
$$
$$
=H(t)U(t)\left|\psi (t)\right>, \eqno(A3)
$$
or equivalently 
$$
i\hbar \left({d\left|\psi (t)\right> \over dt}\right)=
{\cal H}(t)\left|\psi (t)\right> , 
\eqno(A4)
$$
where 
$$
{\cal H}(t)=U^\dagger (t)H(t)U(t)-i\hbar U^\dagger (t)
{dU(t)\over dt}.
\eqno(A5)
$$
Eq.(A5) is the required time dependent canonical transformation.

\bigskip 
\centerline{\bf  MATHEMATICAL APPENDIX B}
\bigskip

From Eqs.(1) and (14) it follows that 
$$
S^\dagger {\bf p}_a S={\bf p}_a+{q_a\over c}{\bf A}.
\eqno(B1)
$$
Eqs.(B1) and (11) then imply 
$$
S^\dagger H_{mol}({\bf A}) S=H_{mol}({\bf A}=0).
\eqno(B2)
$$
Thus, the vector potential in the dipole interaction theory
may be ``gauged away''. However, the dipole interaction returns 
to the full Hamiltonian since Eqs.(9) and (14) imply that 
$$
S^\dagger a_k S=a_k+i({\bf e}_k {\bf \cdot \mu })
\sqrt{2\pi \over \Omega \hbar \omega_k }
\eqno(B2)
$$
and 
$$
S^\dagger a^\dagger _k S=a^\dagger_k-i({\bf e}^*_k {\bf \cdot \mu })
\sqrt{2\pi \over \Omega \hbar \omega_k }\ .
\eqno(B3)
$$
From Eqs.(7), (B2) and (B3) it follows that 
$$
S^\dagger H_{rad}S=H_{rad} - {\bf \mu \cdot E} + W_{mol}
\eqno(B4)
$$ 
where the operator electric field \begin{math} {\bf E}  \end{math} is 
given in Eq.(10), and the molecular self energy  
\begin{math} W_{mol} \end{math} is given in Eq.(17). Thus the same 
canonical transformation which gauges away the vector potential in the 
molecule Hamiltonian, brings back the electric dipole interaction 
\begin{math} -{\bf \mu \cdot E} \end{math} when applied to 
the radiation Hamiltonian via Eq.(B4). Finally, from Eqs.(B2) and (B4) 
it follows that 
$$
S^\dagger \big(H_{rad}+H_{mol}({\bf A})\big)S=
$$
$$
H_{rad}+H_{mol}({\bf A}=0)+W_{mol}-{\bf \mu \cdot E},
\eqno(B5)
$$
and Eqs.(16) and (18) then follows from Eq.(B5). The proof 
of Eq.(18) has been completed.

\bigskip 
\centerline{\bf ACKNOWLEDGMENT}
\bigskip 
We wish to thank Dr. Flavio Fontana for an initial discussion of 
this problem.

\end{document}